\documentclass[11pt,onecolumn]{article}
\usepackage{amsfonts}

\usepackage{amsmath}
\usepackage{amssymb}
\usepackage[top=1in, bottom=1in, left=1.25in, right=1.25in]{geometry}

\usepackage{ifpdf}

\usepackage{cite}
\usepackage[dvips]{graphicx}
\graphicspath{{../eps/}}
\DeclareGraphicsExtensions{.eps}
\usepackage{color, soul} 

\usepackage{array}

\begin{document}
\title{$l_0$ Norm Constraint LMS Algorithm for Sparse System Identification}

\author{Yuantao~Gu\thanks{This work was partially supported by National Natural Science Foundation of China (NSFC 60872087 and NSFC U0835003). The authors are with the Department of Electronic Engineering, Tsinghua University, Beijing 100084, China. The correspond author of this paper is Yuantao Gu (e-mail: gyt@tsinghua.edu.cn).},~
        Jian~Jin,~
        and~Shunliang~Mei}

\date{Received Mar. 24, 2009; accepted June 5, 2009.\\\vspace{1em}
This article appears in \textsl{IEEE Signal Processing Letters}, 16(9):774-777, 2009.}

\maketitle

\begin{abstract}
In order to improve the performance of Least Mean Square (LMS) based
system identification of sparse systems, a new adaptive algorithm is
proposed which utilizes the sparsity property of such systems. A
general approximating approach on $l_0$ norm -- a typical metric  of
system sparsity, is proposed and integrated into the cost function
of the LMS algorithm. This integration is equivalent to add a zero
attractor in the iterations, by which the convergence rate of small
coefficients, that dominate the sparse system, can be effectively
improved. Moreover, using partial updating method, the computational
complexity is reduced. The simulations demonstrate that the proposed
algorithm can effectively improve the performance of LMS-based
identification algorithms on sparse system.

\textbf{Keywords:} adaptive filter, sparsity, $l_0$ norm, Least Mean Square (LMS).
\end{abstract}


\section{Introduction}
%
%
%
%

\begin{figure}
\centering
\includegraphics[width=4in]{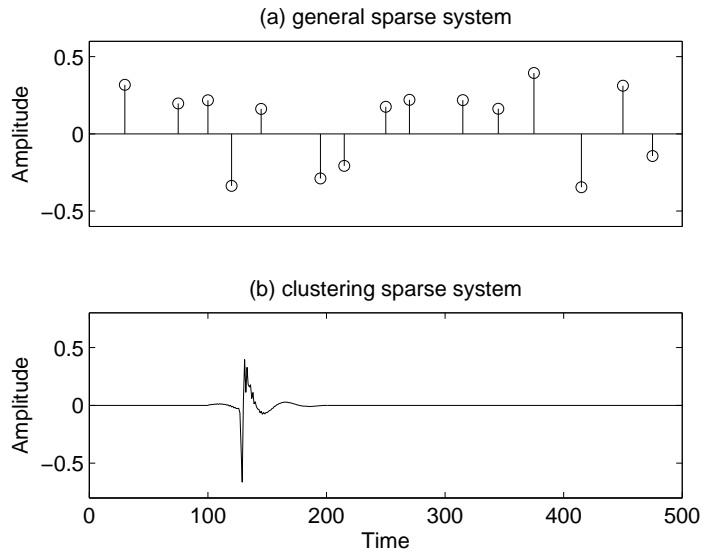}
\caption{Typical sparse system.} \label{sparsec}
\end{figure}

A sparse system is defined whose impulse response
contains many near-zero coefficients and few large ones. Sparse
systems, which exist in many applications, such as Digital TV
transmission channels \cite{Schreiber} and the echo paths
\cite{Duttweiler}, can be further divided to general sparse systems
(Fig.~\ref{sparsec}.~a) and clustering sparse systems
(Fig.~\ref{sparsec}.~b, ITU-T G.168). A clustering sparse system
consists of one or more clusters, wherein a cluster is defined as a
gathering of large coefficients. For example, the acoustic echo path
is a typical single clustering sparse system, while the echo path of
satellite links is a multi-clustering system which includes several
clusters.

There are many adaptive algorithms for system identification, such
as Least Mean Squares (LMS) and Recursive Least Squares (RLS)
\cite{ASP}. However, these algorithms have no particular advantage
in sparse system identification due to no use of sparse
characteristic. In recent decades, some algorithms have exploited
the sparse nature of a system to improve the identification
performance
\cite{Duttweiler,Etter,IPNLMS,Naylor,Li,Nascimento,Exploiting,ribas}.
As far as we know, the first of them is Adaptive Delay Filters (ADF)
\cite{Etter}, which locates and adapts each selected tap-weight
according to its importance. Then, the concept of proportionate
updating was originally introduced for echo cancellation application
by Duttweiler \cite{Duttweiler}. The underlying principle of
Proportionate Normalized LMS (PNLMS) is to adapt each coefficient
with an adaptation gain proportional to its own magnitude. Based on
PNLMS, there exists many improved PNLMS algorithms, such as IPNLMS
\cite{IPNLMS} and IIPNLMS \cite{Naylor}. Besides the above mentioned
algorithms, there are various improved LMS algorithms on clustering
sparse system \cite{Li, Nascimento,ribas}. These algorithms  locate
and track non-zero coefficients by dynamically adjusting the length
of the filter. The convergence behaviors of these algorithms depend
on the span of clusters (the length from the first non-zero
coefficient to the last one in an impulse response). When the span
is long and close to the maximum length of the filter or the system
has multiple clusters, these algorithms have no advantage compared
to the traditional algorithms.

Motivated by Least Absolutely Shrinkage and Selection Operator
(LASSO) \cite{LASSO} and the recent research on Compressive Sensing
(CS) \cite{Donoho}, a new LMS algorithm with $l_0$ norm constraint
is proposed in order to accelerate the sparse system identification.
Specifically, by exerting the constraint to the standard LMS cost
function, the solution will be sparse and the gradient descent
recursion will accelerate the convergence of near-zero coefficients
in the sparse system. Furthermore, using partial updating method,
the additional computational complexity caused by $l_0$ norm
constraint is far reduced. Simulations show that the new algorithm
performs well for the sparse system identification.

\section{New LMS Algorithm}

The estimation error of the adaptive filter output with respect to
the desired signal $d(n)$ is
\begin{equation}
e(n)=d(n)-{\bf x}^{\rm T}(n){\bf w}(n), \label{esterror}
\end{equation}
where ${\bf w}(n) =
\left[w_0(n),w_1(n),\cdots,w_{L-1}(n)\right]^{\rm T}$ and ${\bf
x}(n) = \left[x(n),x(n-1),\cdots,x(n-L+1)\right]^{\rm T}$ denote the
coefficient vector and input vector, respectively, $n$ is the time
instant, and $L$ is the filter length. In traditional LMS the cost
function is defined as squared error $\xi(n)=|e(n)|^2$.

By minimizing the cost function, the filter coefficients are updated
iteratively,
\begin{equation}
w_i(n+1) = w_i(n) + \mu e(n)x(n-i),~~~~\forall 0\leq i < L,
\label{LMSrecu}
\end{equation}
where $\mu$ is the step-size of adaptation.

The research on CS shows that sparsity can be best represented by
$l_0$ norm, in which constraint the sparsest solution is acquired.
This suggests that a $l_0$ norm penalty on the filter coefficients
can be incorporated to the cost function when the unknown parameters
are sparse. The new cost function is defined as
\begin{equation}
\xi(n) = |e(n)|^2 + \gamma\|{\bf w}(n)\|_0, \label{l0LMScost}
\end{equation}
where $\|\cdot\|_0$ denotes $\l_0$ norm that counts the number of
non-zero entries in ${\bf w}(n)$, and $\gamma>0$ is a factor to
balance the new penalty and the estimation error. Considering that
$l_0$ norm minimization is a Non-Polynomial (NP) hard problem, $l_0$
norm is generally approximated by a continuous function. A popular
approximation \cite{Weston} is
\begin{equation}
\|{\bf w}(n)\|_0 \approx \sum_{i=0}^{L-1}\left( 1-{\rm
e}^{-\beta|w_i(n)|} \right), \label{l0app}
\end{equation}
where the two sides are strictly equal when the parameter $\beta$
approaches infinity. According to (\ref{l0app}), the proposed cost
function can be rewritten as
\begin{equation}
\xi(n) = |e(n)|^2 + \gamma\sum_{i=0}^{L-1}\left( 1-{\rm
e}^{-\beta|w_i(n)|} \right). \label{l0LMScost1}
\end{equation}

By minimizing (\ref{l0LMScost1}), the new gradient descent recursion
of filter coefficients is
\begin{equation}
w_i(n+1) = w_i(n) + \mu e(n)x(n-i) -
\kappa\beta \text{sgn}(w_i(n)) {\rm e}^{-\beta|w_i(n)|}, \quad
\forall 0\leq i < L,\label{l0LMSrecu}
\end{equation}
where $\kappa=\mu\gamma$ and sgn($\cdot$) is a component-wise sign
function defined as
\begin{equation}\label{sgnfunc}
    {\text{sgn}}(x)=\left\{
    \begin{array}{cl} \frac{\textstyle x}{\textstyle |x|} & x\neq 0; \\
    0 & {\rm elsewhere}. \end{array} \right.
\end{equation}

To reduce the computational complexity of (\ref{l0LMSrecu}),
especially that caused by the last term, the first order Taylor
series expansions of exponential functions is taken into
consideration,
\begin{equation}\label{taylor}
    {\rm e}^{-\beta |x|}\approx \left\{
    \begin{array}{cl} 1-\beta|x| & |x|\le\frac{\textstyle 1}{\textstyle \beta}; \\
    0 & {\rm elsewhere}. \end{array} \right.
\end{equation}
It is to be noted that the approximation of (\ref{taylor}) is
bounded to be positive because the exponential function is larger
than zero. Thus equation (\ref{l0LMSrecu}) can be approximated as
\begin{equation}
w_i(n+1) = w_i(n) + \mu e(n)x(n-i) + \kappa
f_{\beta}\left(w_i(n)\right)~~~~\forall 0\leq i < L,
\label{l0LMSrecu1}
\end{equation}
where
\begin{equation}\label{falpha}
    f_{\beta}(x)=\left\{
    \begin{array}{cc} \beta^2x+\beta & -\frac{\textstyle 1}{\textstyle \beta}\le x < 0; \\
    \beta^2x-\beta & 0 < x \le \frac{\textstyle 1}{\textstyle \beta}; \\
    0 & {\rm elsewhere}. \end{array} \right.
\end{equation}

The algorithm described by (\ref{l0LMSrecu1}) is denoted as
$l_0$-LMS. Its implementation costs more than traditional LMS due to
the last term in the right side of (\ref{l0LMSrecu1}). It is
necessary, therefore, to reduce the computational complexity
further. Because the value of the last term does not change
significantly during the adaptation, the idea of partial updating
\cite{AFPU}\cite{Godavarti} can be used. Here the simplest method of
sequential LMS is adopted. That is, at each iteration, one in $Q$
coefficients (where $Q$ is a given integer in advance) is updated
with the latest $f_\beta(w_i(n))$, while those calculated in the
previous iterations are used for the other coefficients. Thus, the
excessive computational complexity of the last term is one in $Q$th
of the original method. More detailed discussion on partial update
can be found in \cite{AFPU}. The final algorithm is described using
MATLAB like pseudo-codes in TABLE \ref{Pseudo-codes}.

\begin{table}[t]
\renewcommand{\arraystretch}{1.3}
\caption{The Pseudo-codes of $\l_0$-LMS} \label{Pseudo-codes}
\centering
\begin{tabular}{l}
\hline
Given L, Q, $\mu$, $\beta$, $\kappa$; \\
Initial w = zeros(L,1), f = zeros(L,1);  \\
For i = 1,2,$\cdots$  \\
\hspace{1.5em}input new x and d;  \\
\hspace{1.5em}e = d - x'*w;  \\
\hspace{1.5em}t = mod(i,Q); \\
\hspace{1.5em}f(t+1:Q:L) = -$\beta$*max(0, 1 - $\beta$*abs(w(t+1:Q:L))).*sign(w(t+1:Q:L)); \\
\hspace{1.5em}w = w + $\mu$*e*x + $\kappa$*f; \\
End \\
\hline
\end{tabular}
\end{table}

In addition, the proposed $l_0$ norm constraint can be readily
adopted to improve most LMS variants, e.g. NLMS \cite{ASP}, which
may be more attractive than LMS because of its robustness. The new
recursion of $l_0$-NLMS is
\begin{equation}
w_i(n+1) = w_i(n) + \mu \frac{e(n)x(n-i)}{\delta+{\bf x}^{\rm
T}(n){\bf x}(n)} + \kappa f_{\beta}\left(w_i(n)\right),~~~~\forall
0\leq i < L, \label{l0NLMSrecu}
\end{equation}
where $\delta>0$ is the regularization parameter.

\section{Brief Discussion}
\begin{figure}[t]
\centering
\includegraphics[width=4in]{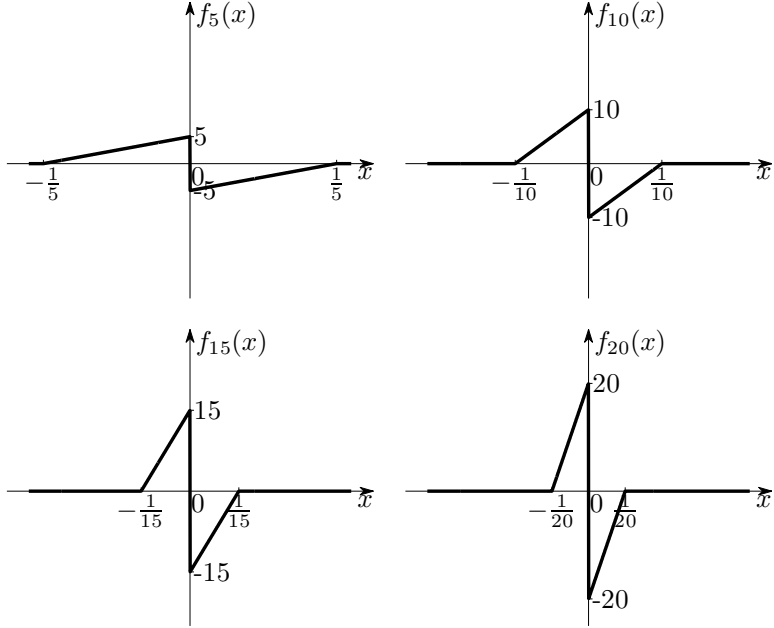}
\caption{The curves of function $f_\beta(x)$ with $\beta$=5, 10, 15,
20.} \label{fx}
\end{figure}

The recursion of filter coefficients in the traditional LMS can be
expressed as
\begin{equation}
{\bf w}_{\rm new} = {\bf w}_{\rm prev} + {\rm gradient~correction},
\end{equation}
where the filter coefficients are updated along the negative
gradient direction. Equation (\ref{l0LMSrecu1}) can be presented in
the similar way,
\begin{equation}
{\bf w}_{\rm new} = {\bf w}_{\rm prev} + {\rm gradient~correction} +
{\rm zero~attraction},
\end{equation}
where zero attraction means the last term in (\ref{l0LMSrecu1}),
$\kappa f_{\beta}\left(w_i(n)\right)$, which imposes an attraction
to zero on small coefficients. Particularly, referring to
Fig.~\ref{fx}, after each iteration, a filter weight will decrease a
little when it is positive, or increase a little when it is
negative. Therefore, it seems that in $\mathbb{R}^{L}$ space of tap
coefficients, \emph{an attractor}, which attracts the non-zero
vectors, exists at the coordinate origin. The range of attraction
depends on the parameter, $\beta$.

The function of \emph{zero attractor} leads to the performance
improvement of $l_0$-LMS in sparse system identification. To be
specific, in the process of adaptation, a tap coefficient closer to
zero indicates a higher possibility of being zero itself in the
impulse response. As shown in Fig.~\ref{fx}, when a coefficient is
within a neighborhood of zero, $(-1/\beta,1/\beta)$, the closer it
is to zero, the greater the attraction intensity is. When a
coefficient is out of the range, no additional attraction is
exerted. Thus, the convergence rate of those near-zero coefficients
will be raised. In conclusion, the acceleration of convergence of
near-zero coefficients will improve the performance of sparse system
identification since those coefficients are in the majority.

According to the above analysis, it can be readily accepted that
$\beta$ and $\kappa$ determine the performance of the proposed
algorithm. Here a brief discussion about the choice of these two
parameters will be given.

\begin{itemize}
  \item The choice of $\beta$: As mentioned above, strong attraction
  intensity or a wide attraction range, which means the tap coefficients
  are attracted more, will accelerate the convergence. According to
  Fig.~\ref{fx}, a large $\beta$ means strong intensity but a narrow
  attraction range. Therefore, it is
  difficult to evaluate the impact of $\beta$ on the convergence
  rate. For practical purposes, Bradley and Mangasarian in \cite{Weston} suggest to set the
  value of $\beta$ to some finite value like $5$ or increased slowly
  throughout the iteration process for better approximation. Here, $\beta=5$ is also
  proper. Further details are omitted here for brevity. And readers of interest please refer to \cite{Weston}.

  \item The choice of $\kappa$: According to (\ref{l0LMScost}) or (\ref{l0LMSrecu1}), the parameter
  $\kappa$ denotes the importance of $l_0$ norm or the intensity of attraction.
  So a large $\kappa$ results in a faster convergence
  since the intensity of attraction increases as $\kappa$ increases.
    On the other hand, steady-state misalignment also increases as
    $\kappa$ increases. After the adaptation reaches steady state, most
  filter weights are near to zero due to the sparsity. We have $|\kappa f_\beta(w_i(n))|\approx\kappa\beta$ for most $i$.
  Regarding to those near-zero coefficients,
  $w_i(n)$ will move randomly in the small neighborhood of zero,
  driven by the attraction term as well as the gradient noise term.
 Therefore, a large $\kappa$ results
  in a large steady-state misalignment. In conclusion, the parameters $\kappa$ are determined by
    the trade-off between adaptation speed and adaptation quality in
    particular applications.
\end{itemize}

\section{Simulations}

The proposed $l_0$-NLMS is compared with the conventional algorithms
NLMS, Stochastic Taps NLMS (STNLMS) \cite{Li}, IPNLMS, and IIPNLMS
in the application of sparse system identification. The effect of
parameters of $l_0$-LMS is also tested in various scenarios.
$\beta=5$ and the proposed partial updating method with $Q$ = $4$
for $l_0$-LMS and $l_0$-NLMS is used in all the simulations.

\begin{figure}[t]
\centering
\includegraphics[width=4in]{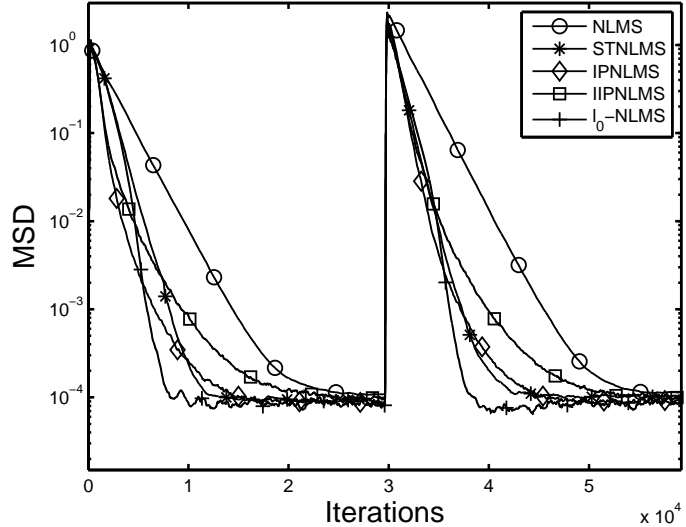}
\caption{Comparison of convergence rate for five different
algorithms, driven by colored signal.} \label{simucor}
\end{figure}

The first experiment is to test the convergence and tracking
performance of the proposed algorithm driven by a colored signal.
The unknown system is a network echo path, which is initialized with
the echo path model 5 in ITU-T recommendation, delayed by 100 taps
and tailed zeros (clustering sparse system, Fig.~\ref{sparsec}.~b).
After $3\times 10^4$ iterations, the delay is enlarged to 300 taps
and the amplitude decrease 6dB. The input signal is generated by
white Gaussian noise $u(n)$ driving a first-order Auto-Regressive
(AR) filter, $x(n)=0.8x(n-1)+u(n)$, and $x(n)$ is normalized. The
observed noise is white Gaussian with variance $10^{-3}$. The five
algorithms are simulated for a hundred times, respectively, with
parameters $L=500, \mu = 1$. The other parameters as follows.
\begin{itemize}
  \item IPNLMS \cite{IPNLMS} and IIPNLMS \cite{Naylor}: $\rho = 10^{-2}, \alpha = 0,
\alpha_1 = -0.5, \alpha_2 =0.5, \Gamma = 0.1$;
  \item ST-NLMS: the initial positions of the first and
last active taps of the primary filter are 0 and 499, respectively;
those of the auxiliary filter are randomly chosen;
  \item $l_0$-NLMS: $\kappa=8\times 10^{-6}$;
\end{itemize}
Please notice that the parameters of all algorithms are chosen to
make their steady-state errors the same. The Mean Square Deviations
(MSDs) between the coefficients of the adaptive filter and the
unknown system are shown in Fig.~\ref{simucor}. According to
Fig.~\ref{simucor}, the proposed $l_0$-NLMS reaches steady-state
first among all algorithms. In addition, when the unknown system
abruptly changes, again the proposed algorithm reaches steady-state
first.

\begin{figure}[t]
\centering
\includegraphics[width=4in]{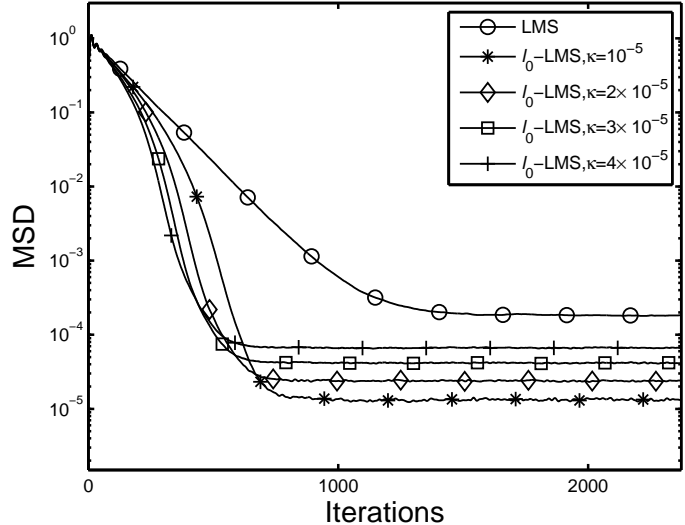}
\caption{Learning curves of $l_0$-LMS with different $\kappa$,
driven by white signal.} \label{simu1vark}
\end{figure}

The second experiment is to test the convergence performance of
$l_0$-LMS with different parameters $\kappa$. Suppose that the
unknown system has $128$ coefficients, in which eight of them are
non-zero ones (their locations and values are randomly selected).
The driven signal and observed noise are white, Gaussian with
variance $1$ and $10^{-4}$, respectively. The filter length is
$L=128$. The step-size of $l_0$-LMS is fixed to $10^{-2}$, while
$\kappa$ is with different values. After a hundred times run, their
MSDs are shown in Fig.~\ref{simu1vark}, in which MSDs of LMS
($\mu=10^{-2}$) are also plotted for reference. It is evidently
recognized that $l_0$ norm constraint algorithm converges faster
than its ancestor. In addition, from the figure, it is obvious that
a larger $\kappa$ results in a higher convergence rate but a larger
steady-state misalignment. These illustrate again that a delicate
compromise should be made between the convergence rate and
steady-state misalignment for the choice of $\kappa$ in practice.
Certainly, the above results are consistent with the discussion in
the previous section.

\begin{figure}[!t]
\centering
\includegraphics[width=4in]{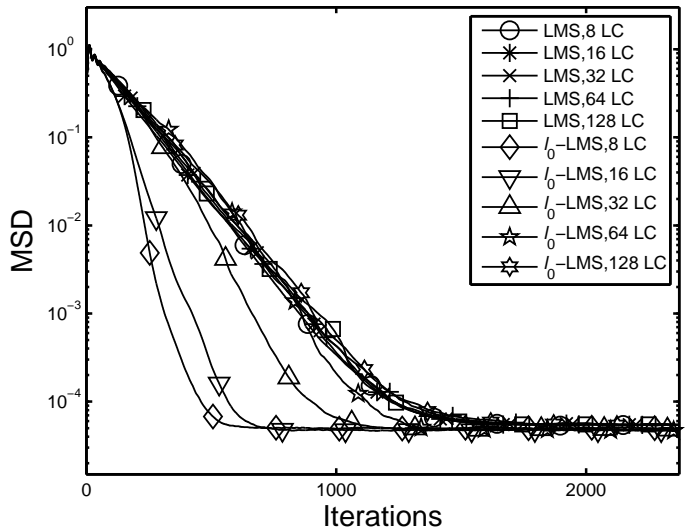}
\caption{Learning curves of $l_0$-LMS and LMS with different
sparsities, driven by white signal, where LC denotes Large
Coefficients.} \label{simu4}
\end{figure}

The third experiment is to test the performance of $l_0$-LMS
algorithm with various sparsities. The unknown system is supposed to
have a total of $128$ coefficients and is a general sparse system.
The number of large coefficients varies from $8$ to $128$, while the
other coefficients are Gaussian noise with a variance of $10^{-4}$.
The input driven signal and observed noise are the same as that in
the first experiment. The filter length is also $L=128$. In order to
compare the convergence rate in all scenarios, the step-sizes are
fixed to $6\times 10^{-3}$. Parameter $\kappa$ is carefully chosen
to make their steady-state error the same (TABLE \ref{simu3param}).
All algorithms are simulated $100$ times respectively and their MSDs
are shown in Fig.~\ref{simu4}. As predicted, the number of the large
coefficients has no influence on the performance of LMS. However,
the convergence rate decreases as the number of large coefficients
increases for $l_0$-LMS. Therefore, the new algorithm is sensitive
to the sparsity of system, that is, a sparser system has better
performance. As the number of large coefficients increases, the
performance of $l_0$-LMS is gradually degraded to that of standard
LMS. Meanwhile, it is to be emphasized that in all cases, the
$l_0$-LMS algorithm is never worse than LMS.

\section{Conclusion}
In order to improve the performance of sparse system identification,
a new LMS algorithm is proposed in this letter by introducing $l_0$
norm, which has vital impact on sparsity, to the cost function as an
additional constraint. Such improvement can evidently accelerate the
convergence of near-zero coefficients in the impulse response of a
sparse system. To reduce the computing complexity, a method of
partial updating coefficients is adopted. Finally, simulations
demonstrate that $l_0$-LMS accelerates the identification of sparse
systems. The effects of algorithm parameters and unknown system
sparsity are also verified in the experiments.

\begin{table}[!t]
\renewcommand{\arraystretch}{1.3}
\caption{The parameters of $l_0$-LMS in the 3rd experiment.}
\label{simu3param} \centering
\begin{tabular}{cccccc}
\hline  LCN \footnotemark{} & 8 & 16 & 32 & 64 & 128 \\
\hline
$\kappa$ & $8\times 10^{-5}$  &  $5.5\times 10^{-5}$   & $4.5\times 10^{-5}$  & $3.5\times 10^{-5}$  & $10^{-6}$     \\
\hline
\end{tabular} \\
\begin{tabular}{l} [1]LCN denotes Large Coefficients Number.
\end{tabular}
\end{table}

\section*{Acknowledgment}
The authors are very grateful to the anonymous reviewers for their
part in improving the quality of this paper.

\end{document}